\documentclass[aps,prd,twocolumn,nofootinbib,amsmath,amssymb,showpacs]{revtex4-1}

\usepackage{graphics}
\usepackage{color}
\usepackage{hyperref}

\usepackage{lipsum}

\begin{document}
 
\title{Nucleating quark droplets in the core of magnetars}

\author{D. {\sc Kroff}}
\email{kroff@if.ufrj.br}
\author{E. S. {\sc Fraga}}
\email{fraga@if.ufrj.br}

\affiliation{Instituto de F\'\i sica, Universidade Federal do Rio de Janeiro,
Caixa Postal 68528, 21941-972, Rio de Janeiro, RJ, Brazil}

\date{\today}


\begin{abstract}
  To assess the possibility of homogeneous nucleation of quark matter in magnetars, we investigate the formation of chirally symmetric droplets in a cold and 
  dense environment in the presence of an external magnetic field. As a framework, we use the one-loop effective potential of the two-flavor quark-meson model. 
  Within the thin-wall approximation, we extract all relevant nucleation parameters and provide an estimate for the typical time scales for the chiral phase 
  conversion in magnetized compact star matter. We show how the critical chemical potential, critical radius, correlation length and surface tension are 
  affected, and how their combination to define the nucleation time seems to allow for nucleation of quark droplets in magnetar matter even for not so small 
  values of the surface tension.
\end{abstract}



\maketitle


\section{Introduction}

  The thermodynamics of strong interactions in cold and dense matter under the influence of strong magnetic fields is of clear relevance in the description of
  magnetars. These objects correspond to a class of compact stars \cite{Glendenning:2000} whose magnetic fields can reach up to $10^{15}$G at the surface 
  \cite{Duncan:1992hi,*Thompson:1993hn}, and even higher yet unknown in the core. (Ref. \cite{Ferrer:2010wz} presents an upper limit of 
  $10^{20}$G $\sim 60 m_{\pi}^{2}$, which could be achieved in the core of self-bound strange stars. This value exceeds even the magnetic fields generated in 
  peripheral heavy ion collisions at high energy \cite{magnetic-HIC} and would certainly affect the phase structure and phase conversion of strong 
  interactions.)

  The full description of the structure and dynamics of formation of these objects depends on the knowledge of the equation of state for the matter they are 
  built of, including possible condensates and new phases that are energetically more favored as baryon density is increased \cite{Heiselberg:1999mq}. In 
  particular, for high enough energy densities, one expects that strongly interacting matter becomes deconfined and essentially chiral \cite{Rischke:2003mt}, 
  so that chiral quark matter could provide the relevant degrees of freedom in the core of compact stars \cite{Weber:2004kj,Alford:2006vz}

  In fact, it was shown that deconfinement can happen at an early stage of a core-collapse supernova process, which could result not only in a delayed explosion
  but also in a neutrino signal of the presence of quark matter in compact stars \cite{Sagert:2008ka}. However, as discussed in Ref. \cite{Mintz:2009ay} (see 
  also \cite{Bombaci:2009jt}), this possibility depends crucially on the time scales of phase conversion. Since one expects a first-order nature for the chiral
  and the deconfinement transitions in cold and dense matter, this process would be guided by bubble nucleation, which is usually slow, or spinodal 
  decomposition, depending on how fast the system reaches the spinodal instability as compared to the nucleation rate. It has been shown in Ref. 
  \cite{Mintz:2009ay} that a key ingredient is the surface tension, which was later estimated in Refs. 
  \cite{Palhares:2010be,Pinto:2012aq,Mintz:2012mz,Lugones:2013ema,Ke:2013wga}.

  The surface tension for magnetized quark matter was estimated within the Nambu--Jona-Lasinio model in Ref. \cite{Garcia:2013eaa}, exhibiting an interesting 
  non-monotonic behavior as a function of the magnetic field. However, as has become clear in the analysis of Ref. \cite{Palhares:2010be}, different ingredients
  in the nucleation process (such as the critical radius, the critical chemical potential and the surface tension) can react very differently to variations of
  an external control parameter. Since the time scales for the phase conversion process are built from a non-trivial combination of these quantities, one needs
  to compute how each of them is affected by an external magnetic field to assess whether nucleation can be the driving mechanism for the chiral transition in
  the case of magnetar matter.

  In this paper we assess the possibility of homogeneous nucleation of quark matter in magnetars by investigating the formation of chirally symmetric droplets
  in a cold and dense environment in the presence of an external magnetic field. As a framework, we use the linear sigma model coupled to quarks, also known as
  the two-flavor quark-meson model. From the one-loop effective potential, and within the thin-wall approximation, we extract all relevant nucleation 
  parameters and provide an estimate for the typical time scales for the chiral phase conversion in magnetized compact star matter. We show how the critical 
  chemical potential, the correlation length, the critical radius, the surface tension and the nucleation rate are affected. The nucleation time is obtained 
  from a non-trivial combination of these quantities and seems to favor nucleation even for not so small values of the surface tension.

  The paper is organized as follows. In Section \ref{sec:effth} we briefly describe the effective model and the approximations used to compute the effective 
  potential. Section \ref{sec:tension} shows how we proceed in order to obtain all nucleation parameters in the thin-wall approximation. Our results for the 
  relevant quantities related to the nucleation process are presented in Section \ref{sec:results}. Section \ref{sec:conclusion} presents our summary.


\section{Effective Theory}
  \label{sec:effth}

  \subsection{General Framework}
  
    To study the phase conversion process, we adopt the linear sigma model coupled to quarks (LSMq) \cite{Scavenius:2000qd} as our effective theory description 
    of the chiral sector of strong interactions. The Lagrangian is given by
      \begin{equation}
	\label{LSMq_lagrangian}
	\begin{split}
	  \mathcal{L} = & \,\bar{\psi}_f [i\gamma^{\mu}\partial_{\mu}-g(\sigma+i\gamma_5\boldsymbol{\tau}\cdot\boldsymbol{\pi})]\psi_f \\
	  & + \frac{1}{2}(\partial_{\mu}\sigma\partial^{\mu}\sigma + \partial_{\mu}\boldsymbol{\pi}\cdot\partial^{\mu}\boldsymbol{\pi}) \\
	  & -\frac{\lambda}{4}(\sigma^2+\boldsymbol{\pi}^2-v^2)^2+h\sigma  \ .
	\end{split}
      \end{equation}

    The model contains a fermionic SU(2) chiral doublet, $\psi_f$, representing the up and down constituent quarks, and four mesons -- one scalar, $\sigma$, 
    and three pseudoscalars, $\boldsymbol{\pi}$. The mesons can be grouped into a single $O(4)$ chiral field $\phi \equiv (\sigma,\boldsymbol{\pi})$. It is 
    well know that the LSMq reproduces correctly all the chiral low energy phenomenology of strong interactions, such as mesons masses and the spontaneous 
    and (small) chiral symmetry breaking, which are present in the mesonic self-interaction potential. The model parameters are fixed accordingly 
    \cite{Scavenius:2000qd}. Moreover, it was argued \cite{Pisarski:1983ms} that both QCD with two flavors of massless quarks and the model we consider belong
    to the same universality class, thus exhibiting the same behavior at criticality\footnote{Recent lattice results seem to challenge this connection in the
    chiral limit \cite{Bonati:2014kpa}, although further detailed studies are still necessary.}.
    
    Due to spontaneous symmetry breaking, the $\sigma$ field acquires a non-vanishing vacuum expectation value. However, for sufficiently high temperatures, 
    the condensate melts and chiral symmetry is approximately restored. Therefore, in this context the expectation value of the $\sigma$ field plays the role 
    of an approximate order parameter for the chiral transition, being exact only in the limit of vanishing quark (and pion) masses, which happens for
    $h = 0$. In this limit, the model becomes truly chiral, and the pions behave as Goldstone bosons. So, to investigate the phase conversion in the LSMq, one
    ultimately needs to study how the expectation value $\langle \sigma \rangle =\bar\sigma$ varies as a function of the relevant control parameters, such as
    temperature, chemical potentials and external fields. As usual in this approach, the effective potential formalism rises as the appropriate means for the
    description of phase transitions. In the spirit of effective theory descriptions, we will not be concerned with numerical precision, but rather in 
    obtaining qualitative information about the system under consideration. Moreover, in order to perform a semi-analytic study, some simplifying 
    approximations are needed. 
    
    The first regards the fermionic contribution to the effective potential. As the action is quadratic in the fermion fields, we can formally integrate over
    the quarks, so that their contribution to the effective potential is given by a determinant. However, as the quarks couple to $\sigma$, one is left to 
    compute a fermionic determinant in the presence of an arbitrary background field, which cannot be done in closed form, unless for systems in $1+1$ 
    dimensions under some special circumstances \cite{AragaoDeCarvalho:1988us,Fraga:1994xd,novikov}. As customary, we consider the quark gas as a thermal bath
    in which the long-wavelength modes of the chiral field evolve, so that the calculation is performed considering a static and homogeneous background field.
    This procedure can be further improved, e.g. via a derivative expansion \cite{Barci:1996ph,Taketani:2006zg,Nickel:2009wj}.
    
    The contribution from the mesons to the effective potential is also subject to simplifying approximations. First, it has been shown that the pions do not 
    affect appreciably the phase conversion process, so their dynamics is usually discarded and the whole analysis can be done setting 
    $\boldsymbol{\pi} = \langle \boldsymbol{\pi} \rangle = 0$. Second, since $\lambda \approx 20$, quantum corrections arising from the sigma self interaction 
    are usually ignored, and its contribution to the effective potential is taken to be classical\footnote{See, however, Ref. \cite{Carter:1996rf}, where the
    authors consider thermal meson fluctuations using resummations, and Ref. \cite{Palhares:2010be}, where the authors compute the one-loop correction to the
    classical potential and treat systematically vacuum terms.}.
     
  \subsection{Effective potential at one loop in a magnetic background}
  
    Our aim is to study the chiral transition in a cold and dense environment in the presence of an external constant and homogeneous magnetic field, as a very
    simplified model for the core of a magnetar. Adapting the previous setup to describe such a system is straightforward. The interaction with the magnetic 
    field is introduced via minimal coupling, i.e the derivatives acting on quarks are traded for $D_{\mu} \equiv \partial_{\mu} + i q A_{\mu}$. Following 
    previous work, we use the aforementioned approximations when computing the effective potential. 
    
    In this setup, the effective potential for two flavors of quarks with $N_c$ colors in the presence of a homogeneous and static magnetic field $\mathbf{B}$ 
    in the cold and dense limit can be written as sum of three contributions \cite{Fraga:2012rr}: 
    \begin{equation}
      \label{eq:effpot1}
      V_{\text{eff}}(\bar{\sigma}) = U_{\text{cl}}(\bar{\sigma}) + U^{\text{vac}}_f(\bar{\sigma},B) + U^{\text{med}}_f(\bar{\sigma},\mu,B) \ .
    \end{equation}

    The first term is just the classical potential for $\sigma$, the second gives the fermionic vacuum contribution 
    \begin{equation}
      \label{eq:ferm_vac}
      \begin{split}
	U^{\text{vac}}_f = -\frac{N_c}{2\pi^2}\sum_f (q_fB)^2 &\Bigg[ \zeta_H'(-1,x_f) \ \ \ + \\
	& -\frac{x^2_f-x_f}{2}\log x_f +\frac{x^2_f}{4}\Bigg],
      \end{split}
    \end{equation}
    where $x_f = M^2_q/(2 |q_f| B)$, $M_q = g\bar{\sigma}$ is the quark dynamically-generated mass, $q_f$ is the electric charge of quark species $f$ and 
    $\zeta_H'$ denotes the derivative with respect to the first argument of the Hurwitz $\zeta$-function. Finally, the last term of Eq. \eqref{eq:effpot1} is
    the medium contribution due to the quarks (see e.g. Ref. \cite{Menezes:2008qt})
    \begin{equation}
      \label{eq:ferm_med}
      \begin{split}
	U^{\text{med}}_f = & -\frac{N_c}{4\pi^2}\sum_f\sum_{\nu=0}^{\nu_{\text{max}}}(2-\delta_{\nu 0})|q_f|B 
	\Bigg[ \mu\sqrt{\mu^2-M^2_{fB}} \ \ \ + \\
	& \left.- M^2_{fB}\log\left(\frac{\mu+\sqrt{\mu^2-M^2_{fB}}}{M_{fB}}\right)\right],
      \end{split}
    \end{equation}
    In this last expression we assume that both fermion species have the same chemical potential $\mu$. In addition, $M^2_{qB} = M^2_q + 2|q_f|B$ denotes the
    magnetic correction to the quark mass and $\nu$ is an integer value that labels Landau levels. The last occupied level is given by:
    \begin{equation}
      \label{eq:numax}
      \nu_{\text{max}} = \left\lfloor \frac{\mu^2-M^2_f}{2|q_f|B} \right \rfloor .
    \end{equation}
    This effective potential exhibits a first-order phase transition for a critical value $\mu_c(B)$ of the chemical potential.
        

\section{Surface tension and nucleation}
  \label{sec:tension}

  Given the effective potential, we can proceed to the study of the phase conversion process driven by the chiral transition. The physical setup we have in 
  mind is that of a collapsing star and, more specifically, the scenario of magnetar formation. Thus, we investigate whether chirally symmetric matter can be
  nucleated as the density increases in the presence of a strong magnetic field.
  
  In our analysis, we focus on homogeneous nucleation. Dynamically, there are two ways by which nucleation can occur: thermal activation and quantum tunneling.
  At the temperatures that correspond to the scenario at hand, of the order of $10 - 30$ MeV, and in the presence of a barrier in the effective potential, 
  thermal activation is by far the dominant way \cite{Mintz:2009ay}. Once the barrier disappears, the initial state of the system is no longer in a metastable
  vacuum, so that spinodal decomposition takes place and the phase conversion is explosive \cite{Gunton:1983}.
  
  It is important to state that there is no contradiction in considering thermal activation of bubbles and taking the cold, i.e. $T\sim 0$, limit to compute
  the effective potential, see Eq. \eqref{eq:ferm_med}. When we focus on thermal nucleation, we are ultimately comparing temperature with the height of barrier
  separating true and false vacua, whereas when we consider the cold limit we compare it with the quark chemical potential. Indeed, in our setup the 
  temperature is high enough to enable thermal activation and low enough to justify the use of the zero-temperature effective potential\footnote{It has been
  shown that thermal fluctuations and quantum vacuum corrections compete when they are included in $V_{\rm eff}$ \cite{Palhares:2010be}.}. 
  
  Our aim is to estimate typical times scales for the nucleation process and to understand under which conditions it is favored. In other words, which are the
  features that can make nucleation happen effectively in magnetar matter, producing chirally symmetric matter in the core of such stars. As  mentioned 
  previously, a key quantity seems to be the surface tension, since it is the amount of energy needed to build up a barrier separating the two phases. In other
  words, the surface tension is the energetic cost to create a bubble.

  \subsection{Extracting nucleation parameters from the effective potential}
  \label{subsec:parameters}
  
    Since we are not concerned with numerical precision, but rather with obtaining reasonable estimates and the qualitative functional behavior, it is 
    convenient to work with approximate analytic relations by fitting the effective potential in the relevant region. This can be done conveniently using a 
    quartic polynomial and imposing the thin-wall limit. In the range between the critical chemical potential, $\mu_c$, and the spinodal, $\mu_{sp}$, the 
    effective potential can be written in the following Landau-Ginzburg form \cite{Scavenius:2000bb,Taketani:2006zg}:
    \begin{equation}
      \label{eq:fit}
      V_{\text{eff}} \approx \sum^4_{n=0} a_n \ \phi^n.
    \end{equation}
    Although this approximation is not able to reproduce the three minima of $V_{\text{eff}}$, the polynomial form gives a good quantitative description of 
    the function in the region containing the two minima representing the symmetric and broken phases as well as the barrier between them.
    
    A quartic potential such as Eq. \eqref{eq:fit} can always be written in the form
    \begin{equation}
	\label{eq:ginzburglandau}
	\mathcal{V}(\varphi) = \alpha(\varphi^2-a^2)^2 + j\varphi ,
    \end{equation}
    with the coefficients above defined in terms of the $a_n$ as follows \cite{Scavenius:2000bb,Taketani:2006zg}:
    \begin{subequations}
	\label{eq:coeff}
	\begin{eqnarray}
	\alpha &=& a_4, \\
	a^2 &=& \frac{1}{2}\left[ -\frac{a_2}{a_4}+\frac{3}{8}\left(\frac{a_3}{a_4}\right)^2\right], \\
	j &=& a_4 \left[ \frac{a1}{a_4}-\frac{1}{2}\frac{a_2}{a_3}+\frac{1}{8}\left(\frac{a_3}{a_4}\right)^3\right], \\
	\varphi &=& \phi +\frac{1}{4}\frac{a_3}{a_4}.
	\end{eqnarray}
    \end{subequations}

    The new potential ${\mathcal V}(\varphi)$ reproduces the original $V_{\rm eff} ( \phi )$ 
    up to a shift in the zero of energy.  We are interested in the effective 
    potential only between $\mu_c$ and $\mu_{sp}$.  At $\mu_c$, we will have two 
    distinct minima of equal depth.  This clearly corresponds to the choice $j = 0$
    in Eq. \eqref{eq:ginzburglandau}, so that ${\mathcal V}$ has minima at $\varphi = \pm a$ and a maximum 
    at $\varphi = 0$.  The minimum at $\varphi = -a$ and the maximum move closer 
    together as the chemical potential is shifted and merge at $\mu_{sp}$.  Thus, 
    the spinodal requires $j/\alpha a^3 = -8/3\sqrt{3}$ in Eq. \eqref{eq:ginzburglandau}. 
    The parameter $j/\alpha a^3$ falls roughly 
    linearly from $0$, at $\mu=\mu_c$, to $-8/3\sqrt{3}$ at the spinodal.
    
    In the thin-wall limit the explicit form of the critical bubble is given by \cite{Fraga:1994xd}
    \begin{equation}
    	\label{eq:crit_bubble}
	\varphi_b(r,\xi,R_c) = \varphi_f + \frac{1}{\xi\sqrt{2\alpha}}\left[1-\tanh{\left(\frac{r-R_c}{\xi}\right)} \right],
    \end{equation}
    where $\varphi_f$ is the new false vacuum, $R_c$ is the radius of the 
    critical bubble, and $\xi=2/m$, with $m^2\equiv {\mathcal V}''(\varphi_f)$, is a measure 
    of the wall thickness.  The thin-wall limit corresponds to $\xi/R_c\ll 1$ \cite{Fraga:1994xd}, 
    which can be rewritten as $(3|j|/8\alpha a^3)\ll 1$.  
    Nevertheless, it was shown in \cite{Scavenius:2000bb,Bessa:2008nw}, for 
    the case of zero density and finite temperature, that the thin-wall limit becomes 
    very inaccurate as one approaches the spinodal. (This is actually a very general 
    feature of this description \cite{Gunton:1983}.) In this vein, 
    the analysis presented below is to be regarded as semi-quantitative and provides estimates, not accurate results.
        
    In terms of the parameters $\alpha$, $a$, and $j$ defined above, one 
    finds \cite{Scavenius:2000bb,Taketani:2006zg}
    \begin{eqnarray}
      \label{eq:par}
      \varphi_{t,f} &\approx& \pm a - \frac{j}{8\alpha a^2}  \quad, \\
      \xi &=& \left[ \frac{1}{\alpha (3\varphi_f^2-a^2)} \right]^{1/2}
      \label{twcorlength}
    \end{eqnarray}
    in the thin-wall limit. The surface tension, $\Sigma$, is given by
    \begin{equation}
      \label{eq:surftension}
      \Sigma\equiv \int_0^{\infty}{\rm d}r~\left( \frac{{\rm d}
      \varphi_b}{{\rm d}r} \right)^2 
      \approx \frac{2}{3\alpha\xi^3} \ ,
    \end{equation}
    and the critical radius is obtained from $R_c = (2\Sigma/\Delta V)$, where
    $\Delta V \equiv V(\phi_f)-V(\phi_t) \approx 2 a | j |$. Finally, the free energy of a 
    critical bubble is given by $F_b=(4\pi\Sigma/3)R_c^2$, and 
    from knowledge of $F_b$ one can evaluate the nucleation rate 
    $\Gamma \sim e^{-F_b/T}$. In calculating thin-wall properties, we shall 
    use the approximate forms for $\phi_t$, $\phi_f$, $\Sigma$, and $\Delta V$ 
    for all values of the potential parameters.


\section{Results}
  \label{sec:results}

  In this section we use the method described above to describe quantitatively the nucleation process in the LSMq in the presence of a magnetic background 
  field. We compute different nucleation parameters for the formation of chirally symmetric droplets in a chirally asymmetric medium for values of the 
  external magnetic field that are compatible with what one  expects to be relevant to magnetar matter. As an initial step, we analyze how the critical 
  chemical potential depends on $B$.
  
  \subsection{Landau level filling and oscillations}

\begin{figure}[ht]
      \begin{center}
	\resizebox*{!}{5.5cm}{\includegraphics{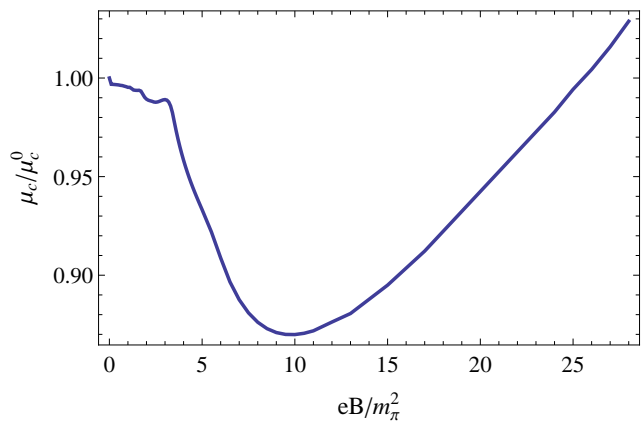}}
      \end{center}
      \caption{\label{fig:mu_c} Critical chemical potential, $\mu_c$ as a function of $B$. The solid line is just a guide for the eye.}
    \end{figure}

    When studying the critical behavior of the LSMq in the presence of an external magnetic field, the first question we should consider is how the position
    of the critical line is affected by $B$. The plot in Fig. \ref{fig:mu_c} shows the behavior of the critical chemical potential, $\mu_c(B)$, normalized by
    the critical chemical potential in the absence of the external field, $\mu_c(0) = \mu^0_c \approx 305$ MeV.

    From the plot it is clear that $\mu_c$ has a nonmonotonic dependence on $B$, it oscillates and reaches a minimum value for $eB \approx 10 m^2_{\pi}$.
    The results show clearly that the presence of a moderate external magnetic field can reduce the value of $\mu_c$ up to 15\%.
    
    The small oscilations observed for $eB \lesssim 4 m^2_{\pi}$ are analogous to the de Haas--van Alphen oscillations in metallic crystals. 
    They are related to the fact that, as we vary the magnetic field, the degeneracy of the Landau levels and the spacing between them are modified, so that 
    the level filling varies with $B$. On the other hand, the behavior for $eB \gtrsim 4 m^2_{\pi}$ is purely due to the lowest Landau level filling.
    For a detailed discussion see Ref. \cite{Schmitt}.
  
  \subsection{Nucleation parameters}
  
    Oscillations are not only seen in the behavior of $\mu_c(B)$. In fact, as the following plots show, all the nucleation parameters have a non-trivial 
    oscillatory dependence on the magnetic field.
    \begin{figure}[ht]
      \begin{center}
	\resizebox*{!}{5.5cm}{\includegraphics{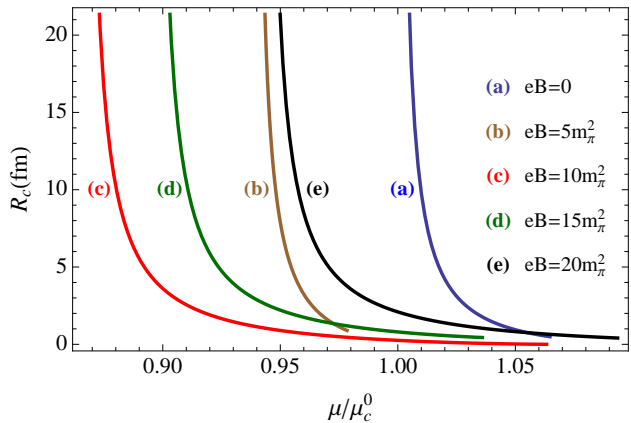}}
      \end{center}
      \caption{\label{fig:Rc} Critical radius of chirally symmetric droplets as a function of the quark chemical potential for different values of $eB$.}
    \end{figure}
    
    Recall that whenever a bubble is formed, its interior tends do lower the free energy of the system, since the field within sits on the true vacuum. On 
    the other hand, the surface of the bubble tends to increase it, as discussed previously. The critical bubble is the one whose energetic gain due to the 
    volume exactly compensates the cost of the surface. Thus, to minimize the energy, any bubble smaller than the critical will shrink and the ones that are
    bigger will expand. Therefore, the radius of the critical bubble, or critical radius, sets the threshold between suppressed and favored bubbles. 

    \begin{figure}[ht]
      \begin{center}
	\resizebox*{!}{5.5cm}{\includegraphics{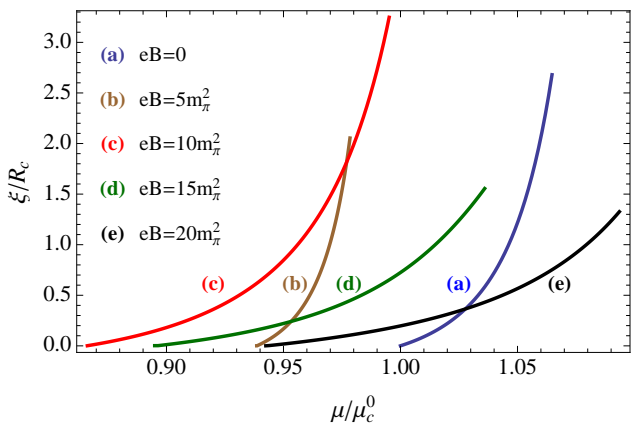}}
      \end{center}
      \caption{\label{fig:xi_o_R} Ratio between the correlation length $\xi$ and the critical radius as a function of quark chemical potential for different 
	      values of $eB$.}
    \end{figure}

    In Fig. \ref{fig:Rc}, we show this quantity as a function of the quark chemical potential for different values of magnetic field. It is interesting to 
    notice that, as a consequence of the critical chemical potential oscillation, the metastable region shifts when the magnetic field varies: first to lower
    values of $\mu$ and then in the opposite direction.
    
    As mentioned in the previous section, the correlation length, $\xi$, provides a measure of the thickness of the bubble wall. The thin-wall approximation
    relies on the assumption that $\xi/R_c \ll 1$ or, equivalently, that the free energy difference between both vacua is small compared to the barrier 
    between them. In Fig. \ref{fig:xi_o_R} we plot this quantity as a function of quark chemical potential. As one should expect, this assumption is reasonable
    far from the spinodal, in the vicinity of the critical line. Nevertheless, in the spirit of providing estimates and the qualitative behavior, we apply 
    the thin-wall limit in the whole range of chemical potentials between $\mu_c$ and $\mu_{sp}$.

    \begin{figure}[ht]
      \begin{center}
	\resizebox*{!}{5.5cm}{\includegraphics{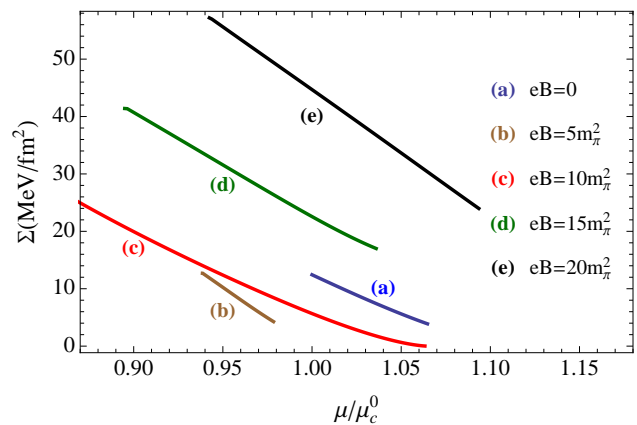}}
      \end{center}
      \caption{\label{fig:sigma} Surface tension as a function of quark chemical potential for different values of $eB$.}
    \end{figure}

    Finally, in Fig. \ref{fig:sigma} we present the results for the surface tension as a function of quark chemical potential for different magnetic fields.
    This plot shows clearly that for $B \lesssim 5 m^2_{\pi}$ the presence of an external magnetic field can actually reduce the energetic cost to build up 
    the bubble wall, which would in principle favor nucleation in this scenario. However, the behavior of $\mu_{c}$ as a function of the magnetic field 
    already gives a hint that the situation is not so straightforward.
    
  \subsection{Estimating typical time scales}
  
    To obtain an estimate of the typical time scales involved in the nucleation of chirally symmetric matter in a cold and dense medium under the influence
    of an external magnetic field, we need first an estimate of the nucleation rate per unit volume, which can be written as $\Gamma \sim T^4_fe^{-F_b/T_f}$,
    where $F_{b}$ is the free energy of the critical bubble and the pre-factor just gives an upper limit with the correct dimensions \cite{Gunton:1983}. 
    Here, we take $T_f = 30$ MeV as a typical temperature for protostars. In doing so we are neglecting the temperature dependence of the critical-bubble 
    free energy or, as we discussed before, using the cold and dense effective potential since the difference scales justify this procedure. 

\begin{figure}[ht]
      \begin{center}
	\resizebox*{!}{5.5cm}{\includegraphics{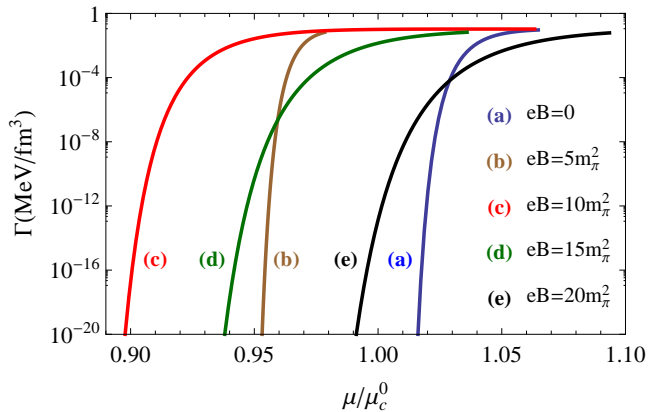}}
      \end{center}
      \caption{\label{fig:gamma} Nucleation rate as a function of the quark chemical potential for different values of $eB$.}
    \end{figure}

    In Fig. \ref{fig:gamma} we show the results for $\Gamma$ as a function of the chemical potential for the same values of magnetic field adopted before. 
    Again a nontrivial oscillation with the magnetic field can be detected.
    
\begin{figure}[ht]
    \begin{center}
      \resizebox*{!}{5.5cm}{\includegraphics{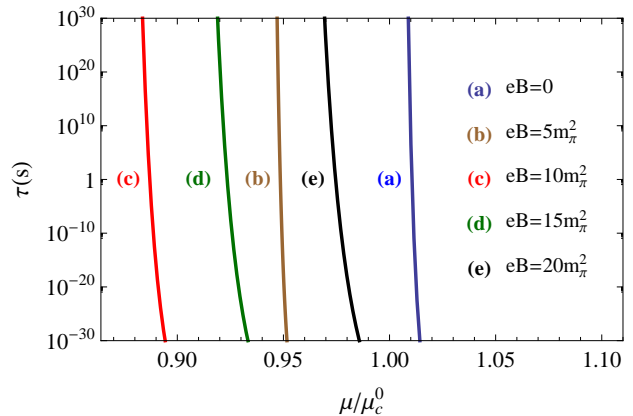}}
    \end{center}
    \caption{\label{fig:tau} Nucleation time as a function of quark chemical potential for different values of $eB$.}
  \end{figure}

    In order to estimate the typical time scales for the phase conversion process, i.e. the formation of chiral quark matter in the core of magnetars, we 
    follow Ref. \cite{Mintz:2009ay} and define the nucleation time as being the time it takes for the nucleation of a single critical bubble inside a volume 
    of 1 km$^3$, which is typical of the core of a proto-neutron star, i.e.:
    \begin{equation}
      \tau \equiv \left( \frac{1}{1 \rm km^3} \right)\frac{1}{\Gamma}.
    \end{equation}
    Fig. \ref{fig:tau} exhibits this quantity as function of the chemical potential for different values of $eB$. The relevant time scale to compare is the 
    time interval the system  takes from the critical chemical potential to the spinodal during the star collapse. Implicitly, in the expression above we are 
    using an approximation of constant density and temperature over the core, which should give a good estimate as the density profile in this region of the 
    star is quite flat \cite{Glendenning:2000}. The plot shows that moderate magnetic fields, $B \lesssim 20 \ m^2_{\pi}$, can actually favor nucleation, as
    a given nucleation time is achieved for lower values of chemical potential.
    

\section{Summary and final remarks}
  \label{sec:conclusion}
  
  In this paper we have used the LSMq minimally coupled to an external classical magnetic field in a cold and dense environment as a simple model to describe 
  critical properties of strongly interacting matter in the core of a magnetar, in particular the likelihood of nucleating approximately chiral quark droplets.
  Using the one-loop effective potential we computed all relevant nucleation parameters within the thin-wall approximation and obtained an estimate for the 
  typical time scales. Our findings indicate that nucleation may be present in the phenomenological interesting range of magnetic fields. Of course, one has 
  also to simulate in detail the evolution of the density profile of the protostar to make any stronger assertion.
    
  The results obtained for the surface tension and nucleation time are very interesting, showing that many different effects sum up in a nontrivial fashion 
  yielding a small nucleation time for cases whose surface tension are not so small. Specifically, the $B$ dependence of $\mu_c$ and the fact that the 
  difference between the free energy of the vacua increases faster for higher values of magnetic field can combine in such a way that cases with a higher 
  surface tension could have a smaller critical radius, ultimately favoring the nucleation picture. Therefore, for magnetars it is not enough to consider the
  behavior (and value) of the surface tension to address the competition between relevant time scales.

  Despite its content of quarks and mesons, the linear sigma model provides essentially a chiral description, i.e. it does not contain essential ingredients
  to describe nuclear matter, such as the saturation density and the binding energy. Nevertheless, this analysis has unveiled how the process of Landau level
  filling affects the nucleation parameters in a nontrivial way, bringing new forms of competition between them and affecting qualitatively the dynamics of
  quark matter formation in compact stars.
  

\begin{acknowledgments}
  The authors would like to thank A. Schmitt for discussion. This work was partially supported by CAPES, CNPq and FAPERJ.
\end{acknowledgments}


\end{document}